\documentclass[a4paper]{article}
\usepackage{amsmath}
\usepackage{geometry}
\usepackage[dvipdfmx]{graphicx}

\begin{document}

\title{Nano and viscoelastic Beck's column on elastic foundation}
\author{Teodor M. Atanackovic\thanks{%
Department of Mechanics, Faculty of Technical Sciences, University of Novi
Sad, Trg D. Obradovica 6, 21000 Novi Sad, Serbia, atanackovic@uns.ac.rs},
Yanni Bouras\thanks{%
Victoria University, College of Engineering and Science, Footscray Park
Campus, Melbourne, Victoria, Australia, yanni.bouras@live.vu.edu.au}, Dusan
Zorica\thanks{%
Mathematical Institute, Serbian Academy of Arts and Sciences, Kneza Mihaila
36, 11000 Belgrade, Serbia, dusan\textunderscore zorica@mi.sanu.ac.rs}}
\maketitle

\begin{abstract}
Beck's type column on Winkler type foundation is the subject of the present
analysis. Instead of the Bernoulli-Euler model describing the rod, two
generalized models will be adopted: Eringen non-local model corresponding to
nano-rods and viscoelastic model of fractional Kelvin-Voigt type. The
analysis shows that for nano-rod, the Herrmann-Smith paradox holds while for
viscoelastic rod it does not.

\end{abstract}

\section{Introduction}

A cantilevered Bernoulli-Euler column subject to a follower force of
constant intensity at the free end, known as Beck's column, represents a
benchmark example of column stability analysis for nonconservative loading,
see \cite{a-ster,Beck,Elishakoff-2005}. Herrmann and Smith analyzed the
problem of dynamic stability for Beck's column when positioned on Winkler
foundation, see \cite{HerSmi}. The critical load causing dynamic instability
(flutter) is found to be independent of the foundation properties. This
phenomenon is known as the Herrmann-Smith paradox. This paradox is aimed to
be resolved in the present analysis by adopting non-local and viscoelastic
constitutive equations as opposed to the classical Bernoulli-Euler relation.

Many authors have been inspired to the study of the Herrmann-Smith paradox
with the intention of removing it. In a number of attempts, the constitutive
equation for the foundation-rod interaction has been changed. The Winkler
model was replaced by viscoelastic models including Kelvin-Voigt, Maxwell
and Zener in \cite{Morgan} and by the fractional Zener model in \cite{A-S-04}%
. It was found that the change of foundation models did not resolve the
paradox. Another attempt included the use of a partial following force in
addition to introducing variable order foundation stiffness, see \cite%
{KirSey}. The analysis showed that these assumptions imply that the critical
force might depend on foundation properties thus resolving the paradox.

In terms of the moment-curvature constitutive equation, if the column is
assumed to be viscoelastic, rather than elastic, the paradox has been shown
to be removed. We refer to \cite{Becker,Katsikadelis}, where viscoelastic
moment-curvature constitutive relationship is adopted in addition to other
generalizations. On resolution of the Herrmann-Smith paradox, another
paradox arises known as the Ziegler (destabilization) paradox, see \cite{Zie}%
. Consider a moment-curvature viscoelastic model that reduces to
Bernoulli-Euler model when the model parameter approaches zero. Then for an
arbitrary small model parameter, the critical force causing dynamic
instability is less then the critical force for the elastic case. This
contradicts the intuitive assumption that dissipation generally increases
the stability regions of mechanical systems thus defining the Ziegler
paradox. For a review and references on paradoxes and errors regarding
stability and vibrations of elastic systems, we refer to \cite{Panovko}.
Dynamic stability problems of viscoelastic Beck's columns were treated in
\cite{Bolotin}. Similar analysis of columns subject to follower force was
conducted in \cite{LanSug}.

In this work we show that for non-local Beck's column, the value of critical
load decreases with the increase of the non-locality parameter. However, it
still does not depend on the foundation properties. Thus the Herrmann-Smith
paradox remains when introducing non-local moment-curvature constitutive
equation. The Herrmann-Smith paradox is removed for the fractional
viscoelastic Beck's column i.e. if the fractional Kelvin-Voigt model is
adopted as a constitutive moment-curvature relation. The fractional
Kelvin-Voigt model reduces to the Bernoulli-Euler model when the order of
fractional differentiation tends to zero. The destabilization paradox was
found to remain for arbitrary small orders of fractional derivative as well.

\section{Problem formulation}

\begin{figure}[h]
\centering
\includegraphics[scale=0.8]{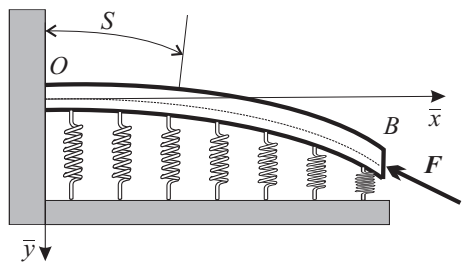}
\caption{Coordinate system and load configuration.}
\label{fig1}
\end{figure}

Let $\bar{x}$ and $\bar{y}$ represent the axes of a rectangular Cartesian
coordinate system with the column in undeformed state being positioned along
the $\bar{x}$-axis, so that its clamped end is in the origin of the
coordinate system, see Figure \ref{fig1}.
System of equations describing the lateral motion in $\bar{x}-\bar{y}$ plane
for Beck's column placed on Winkler foundation consists of: equations of
motion%
\begin{gather}
\frac{\partial }{\partial S}H\left( S,t\right) =\rho \frac{\partial ^{2}}{%
\partial t^{2}}x\left( S,t\right) ,\;\;\;\;\frac{\partial }{\partial S}%
V\left( S,t\right) +q_{y}\left( S,t\right) =\rho \frac{\partial ^{2}}{%
\partial t^{2}}y\left( S,t\right) ,  \label{bek-1} \\
\frac{\partial }{\partial S}M\left( S,t\right) +V\left( S,t\right) \cos
\left( \theta \left( S,t\right) \right) -H\left( S,t\right) \sin \left(
\theta \left( S,t\right) \right) =0,  \label{bek-2}
\end{gather}%
geometrical relations%
\begin{equation}
\frac{\partial }{\partial S}x\left( S,t\right) =\cos \left( \theta \left(
S,t\right) \right) ,\;\;\;\;\frac{\partial }{\partial S}y\left( S,t\right)
=\sin \left( \theta \left( S,t\right) \right) ,  \label{bek-3}
\end{equation}%
and constitutive equations:%
\begin{equation}
M\left( S,t\right) =EI\frac{\partial }{\partial S}\theta \left( S,t\right)
,\;\;\;\;q_{y}\left( S,t\right) =-ky\left( S,t\right) ,  \label{bek-4}
\end{equation}%
for moment-curvature relation (Bernoulli-Euler) and foundation-rod
interaction (Winkler) respectively. The boundary conditions for the system (%
\ref{bek-1}) - (\ref{bek-4}) are%
\begin{gather}
x\left( 0,t\right) =0,\;\;y\left( 0,t\right) =0,\;\;\theta \left( 0,t\right)
=0,  \notag \\
H\left( L,t\right) =-F\cos \theta \left( L,t\right) ,\;\;V\left( L,t\right)
=-F\sin \theta \left( L,t\right) ,\;\;M\left( L,t\right) =0.  \label{Beck 5}
\end{gather}%
In (\ref{bek-1}) - (\ref{bek-4}) time is denoted by $t>0,$ the arc-length of
rod's axis is denoted by $S\in \left[ 0,L\right] $, where $L$ is the length
of the rod, $x$ and $y$ denote the coordinates of an arbitrary point on
rod's axis in the deformed state and $\theta $ is the angle between rod's
axis in deformed and undeformed state. Projections of the contact forces on $%
\bar{x}$ and $\bar{y}$ axes are denoted by $H$ and $V$ respectively, $M$ is
the bending moment, $q_{y}$ denotes the distributed forces per unit length
describing foundation-rod interaction and $F$ is the intensity of the
follower force. Line density, modulus of elasticity, second moment of
inertia of the rod and the stiffness of foundation are denoted by $\rho ,$ $%
E,$ $I$ and $k$ respectively. Note that (\ref{bek-3}) implies that the
column axis is inextensible.

The problem of lateral motion of Beck's column will be treated for the
non-local and viscoelastic constitutive moment-curvature equations as
opposed to the classical Bernoulli-Euler one. For the case when the material
of the rod is modelled by non-local theory of Eringen type, see \cite%
{Elishakoff,eringen,Reddy}, the constitutive equation for bending moment
reads%
\begin{equation}
M\left( S,t\right) -l^{2}\frac{\partial ^{2}}{\partial S^{2}}M\left(
S,t\right) =EI\frac{\partial }{\partial S}\theta \left( S,t\right) ,
\label{nonlocal}
\end{equation}%
where $l$ is the (constant) length scale parameter. Constitutive equation (%
\ref{nonlocal}) is often used when modelling materials with size dependent
properties as in nano-rod theory. Examples include buckling/post-buckling,
vibration and rotation analysis as in \cite%
{AZ-1,Ch-Wa,Thai,Wa-Xi-Kiti,Wa-Zh-1}, \cite{Aranda,LAF,L-L-Y-Z,Wa-Zh} and
\cite{ReddyBorgi}, respectively. Optimization of such rods have also been
studied in \cite{ANV,ANV-1,gsa}.

The constitutive equation%
\begin{equation}
M\left( S,t\right) =EI\left( 1+a\,{}_{0}\mathrm{D}_{t}^{\alpha }\right)
\frac{\partial }{\partial S}\theta \left( S,t\right) ,  \label{FKV}
\end{equation}%
relates to the viscoelastic rod of the fractional Kelvin-Voigt type. In (\ref%
{FKV}), ${}_{0}\mathrm{D}_{t}^{\alpha }$ is the operator of the
Riemann-Liouville fractional derivative of order $\alpha \in \left(
0,1\right) $ given in the form as
\begin{equation*}
{}_{0}\mathrm{D}_{t}^{\alpha }f\left( t\right) =\frac{1}{\Gamma (1-\alpha )}%
\frac{\mathrm{d}}{\mathrm{d}t}\int_{0}^{t}\frac{f(\tau )}{\left( t-\tau
\right) ^{\alpha }}\mathrm{d}\tau ,
\end{equation*}%
see \cite{TAFDE}, where $\Gamma $ is the Euler gamma function and $a$ is the
(constant) model parameter. A number of problems treating lateral vibrations
of viscoelastic rods of fractional type are reviewed in \cite{APSZ-1}.

The trivial solution to system (\ref{bek-1}) - (\ref{Beck 5}) corresponding
to the case when the rod remains straight is independent of the choice of
constitutive equations (\ref{nonlocal}) or (\ref{FKV}) and reads%
\begin{equation*}
x^{0}\left( S,t\right) =S,\;\;y^{0}\left( S,t\right) =0,\;\;\theta
^{0}\left( S,t\right) =0,\;\;H^{0}\left( S,t\right) =-F,\;\;V^{0}\left(
S,t\right) =0,\;\;M^{0}\left( S,t\right) =0.
\end{equation*}%
Assuming $x=x^{0}+\Delta x,\ldots ,$ $M=M^{0}+\Delta M$, where $\Delta
x,\ldots ,\Delta M$ denote the perturbations and upon substitution of
perturbed quantities in (\ref{bek-1}) - (\ref{bek-3}), (\ref{bek-4})$_{2}$
and (\ref{Beck 5}), we obtain $\Delta H=0$ and $\Delta x=0$ as well as%
\begin{gather}
\frac{\partial }{\partial S}\Delta V\left( S,t\right) =k\Delta y\left(
S,t\right) +\rho \frac{\partial ^{2}}{\partial t^{2}}\Delta y\left(
S,t\right) ,\;\;\frac{\partial }{\partial S}\Delta M\left( S,t\right)
+\Delta V\left( S,t\right) +F\Delta \theta \left( S,t\right) =0,
\label{rod-delta 1} \\
\frac{\partial }{\partial S}\Delta y(S,t)=\Delta \theta \left( S,t\right) .
\label{rod-delta 2}
\end{gather}%
Similarly, constitutive moment-curvature relations (\ref{nonlocal}) and (\ref%
{FKV}) become%
\begin{gather}
\Delta M\left( S,t\right) -l^{2}\frac{\partial ^{2}}{\partial S^{2}}\Delta
M\left( S,t\right) =EI\frac{\partial }{\partial S}\Delta \theta \left(
S,t\right) ,  \label{rod-nonlocal} \\
\Delta M\left( S,t\right) =EI\left( 1+a\,{}_{0}\mathrm{D}_{t}^{\alpha
}\right) \left( \frac{\partial }{\partial S}\Delta \theta \left( S,t\right)
\right) ,  \label{rod-k-v}
\end{gather}%
with boundary conditions (\ref{Beck 5}) yielding%
\begin{equation}
\Delta y(0,t)=0,\;\;\Delta \theta (0,t)=0,\;\;\Delta V(L,t)=-F\Delta \theta
\left( L,t\right) ,\;\;\Delta M(L,t)=0.  \label{rod-delta 3}
\end{equation}%
Initial conditions
\begin{equation}
\Delta y\left( S,0\right) =\Delta y_{0}\left( S\right) ,\;\;\frac{\partial }{%
\partial t}\Delta y\left( S,0\right) =\Delta y_{1}\left( S\right) ,
\label{rod-delta-ic}
\end{equation}%
are adjoined to systems (\ref{rod-delta 1}), (\ref{rod-delta 2}), (\ref%
{rod-nonlocal}), (\ref{rod-delta 3}) and (\ref{rod-delta 1}), (\ref%
{rod-delta 2}), (\ref{rod-k-v}), (\ref{rod-delta 3}).

Introducing the dimensionless quantities%
\begin{gather*}
\xi =\frac{S}{L},\;\;\bar{t}=t\sqrt{\frac{EI}{\rho L^{4}}},\;\;y=\frac{%
\Delta y}{L},\;\;y_{0}=\frac{\Delta y_{0}}{L},\;\;y_{1}=\frac{\Delta y_{1}}{L%
}\sqrt{\frac{\rho L^{4}}{EI}},\;\;\vartheta =\Delta \theta , \\
v=\frac{\Delta V\,L^{2}}{EI},\;\;m=\frac{\Delta M\,L}{EI},\;\;\lambda =\frac{%
FL^{2}}{EI},\;\;\bar{k}=\frac{kL^{4}}{EI},\;\;\kappa =\frac{l}{L},\;\;\bar{a}%
=a\left( \sqrt{\frac{EI}{\rho L^{4}}}\right) ^{\alpha }
\end{gather*}%
and upon substitution into (\ref{rod-delta 1}) - (\ref{rod-delta-ic}), after
omitting the bar ($\bar{t}\rightarrow t,$ $\bar{k}\rightarrow k,$ $\bar{a}%
\rightarrow a$), equations of motion (\ref{rod-delta 1}) and geometrical
relation (\ref{rod-delta 2}) read%
\begin{gather}
\frac{\partial }{\partial \xi }v\left( \xi ,t\right) =ky\left( \xi ,t\right)
+\frac{\partial ^{2}}{\partial t^{2}}y\left( \xi ,t\right) ,\;\;\frac{%
\partial }{\partial \xi }m\left( \xi ,t\right) +v\left( \xi ,t\right)
+\lambda \vartheta \left( \xi ,t\right) =0,  \label{DL1} \\
\frac{\partial }{\partial \xi }y(\xi ,t)=\vartheta \left( \xi ,t\right) ,
\label{DL2}
\end{gather}%
with non-local (\ref{rod-nonlocal}) and viscoelastic (\ref{rod-k-v})
constitutive equations becoming%
\begin{gather}
m\left( \xi ,t\right) -\kappa ^{2}\frac{\partial ^{2}}{\partial \xi ^{2}}%
m\left( \xi ,t\right) =\frac{\partial }{\partial \xi }\vartheta \left( \xi
,t\right) ,  \label{DL3} \\
m\left( \xi ,t\right) =\left( 1+a\,{}_{0}\mathrm{D}_{t}^{\alpha }\right)
\left( \frac{\partial }{\partial \xi }\vartheta \left( \xi ,t\right) \right)
,  \label{DL4}
\end{gather}%
and boundary (\ref{rod-delta 3}) and initial conditions (\ref{rod-delta-ic})
transforming to%
\begin{gather}
y(0,t)=0,\;\;\vartheta (0,t)=0,\;\;v(1,t)=-\lambda \vartheta \left(
1,t\right) ,\;\;m(1,t)=0,  \label{DL5} \\
y\left( \xi ,0\right) =y_{0}\left( \xi \right) ,\;\;\frac{\partial }{%
\partial t}y\left( \xi ,0\right) =y_{1}\left( \xi \right) .  \label{DL6}
\end{gather}

\section{Dynamic stability analysis for non-local rod \label{nonlocalsec}}

The non-local constitutive moment-curvature equation will now be adopted in
order to analyse dynamic stability of Beck's column on Winkler foundation
and to determine if the Herrmann-Smith paradox is removed. Adjoining (\ref%
{DL1}), (\ref{DL2}), (\ref{DL3}) it is derived that
\begin{equation}
\left( 1-\kappa ^{2}\lambda \right) \frac{\partial ^{4}}{\partial \xi ^{4}}%
y\left( \xi ,t\right) +\left( \lambda -k\kappa ^{2}\right) \frac{\partial
^{2}}{\partial \xi ^{2}}y\left( \xi ,t\right) -\kappa ^{2}\frac{\partial ^{4}%
}{\partial \xi ^{2}\partial t^{2}}y\left( \xi ,t\right) +\frac{\partial ^{2}%
}{\partial t^{2}}y\left( \xi ,t\right) +ky\left( \xi ,t\right) =0,
\label{eqNonlocal2}
\end{equation}%
subject to boundary (\ref{DL5}) and initial (\ref{DL6}) conditions
\begin{gather}
y(0,t)=0,\;\;\frac{\partial }{\partial \xi }y(0,t)=0,  \label{bc-nl1} \\
\left( 1-\kappa ^{2}\lambda \right) \frac{\partial ^{2}}{\partial \xi ^{2}}%
y\left( 1,t\right) -\kappa ^{2}\frac{\partial ^{2}}{\partial t^{2}}y\left(
1,t\right) -k\kappa ^{2}y\left( 1,t\right) =0,  \label{bc-nl2} \\
\left( 1-\kappa ^{2}\lambda \right) \frac{\partial ^{3}}{\partial \xi ^{3}}%
y\left( 1,t\right) -\kappa ^{2}\frac{\partial ^{3}}{\partial \xi \partial
t^{2}}y\left( 1,t\right) -k\kappa ^{2}\frac{\partial }{\partial \xi }y\left(
1,t\right) =0,  \label{bc-nl3} \\
y\left( \xi ,0\right) =y_{0}\left( \xi \right) ,\;\;\frac{\partial }{%
\partial t}y\left( \xi ,0\right) =y_{1}\left( \xi \right) .  \notag
\end{gather}

Assuming variables can be separated in the form of%
\begin{equation*}
y\left( \xi ,t\right) =U\left( \xi \right) V\left( t\right)
\end{equation*}%
equation (\ref{eqNonlocal2}) becomes%
\begin{equation}
\left( 1-\kappa ^{2}\lambda \right) \frac{U^{\prime \prime \prime \prime
}\left( \xi \right) }{U\left( \xi \right) }+\left( \lambda -k\kappa
^{2}\right) \frac{U^{\prime \prime }\left( \xi \right) }{U\left( \xi \right)
}-\kappa ^{2}\frac{U^{\prime \prime }\left( \xi \right) }{U\left( \xi
\right) }\frac{\ddot{V}(t)}{V(t)}+\frac{\ddot{V}(t)}{V(t)}+k=0,
\label{eqnonlocal3}
\end{equation}%
with boundary conditions (\ref{bc-nl1}) - (\ref{bc-nl3})
\begin{gather}
U(0)=0,\;\;U^{\prime }(0)=0,  \label{bc-nonl1} \\
\left( 1-\kappa ^{2}\lambda \right) \frac{U^{\prime \prime }\left( 1\right)
}{U\left( 1\right) }-\kappa ^{2}\frac{\ddot{V}(t)}{V(t)}-k\kappa ^{2}=0,
\label{bc-nonl2} \\
\left( 1-\kappa ^{2}\lambda \right) \frac{U^{\prime \prime \prime }\left(
1\right) }{U\left( 1\right) }-\kappa ^{2}\frac{U^{\prime }\left( 1\right) }{%
U\left( 1\right) }\frac{\ddot{V}(t)}{V(t)}-k\kappa ^{2}\frac{U^{\prime
}\left( 1\right) }{U\left( 1\right) }=0,  \label{bc-nonl3}
\end{gather}%
where $\left( \cdot \right) ^{\prime }=\frac{\mathrm{d}}{\mathrm{d}\xi }%
\left( \cdot \right) $ and $\left( \cdot \right) ^{\cdot }=\frac{\mathrm{d}}{%
\mathrm{d}t}\left( \cdot \right) .$ Introducing new parameter $\Omega $ so
that%
\begin{equation*}
\ddot{V}(t)+\Omega ^{2}V(t)=0,
\end{equation*}%
equation (\ref{eqnonlocal3}) and boundary conditions (\ref{bc-nonl1}) - (\ref%
{bc-nonl3}) thus become
\begin{gather}
U^{\prime \prime \prime \prime }\left( \xi \right) +r_{1}U^{\prime \prime
}\left( \xi \right) -r_{2}U\left( \xi \right) =0,  \label{eqnonlocal4} \\
U(0)=0,\;\;U^{\prime }(0)=0,  \label{bc-nonloc1} \\
U^{\prime \prime }\left( 1\right) +\kappa ^{2}r_{2}U\left( 1\right)
=0,\;\;U^{\prime \prime \prime }\left( 1\right) +\kappa ^{2}r_{2}U^{\prime
}\left( 1\right) =0,  \label{bc-nonloc2}
\end{gather}%
where
\begin{equation}
r_{1}=\frac{\lambda +\kappa ^{2}\left( \Omega ^{2}-k\right) }{1-\kappa
^{2}\lambda },\;\;r_{2}=\frac{\Omega ^{2}-k}{1-\kappa ^{2}\lambda }.
\label{rs}
\end{equation}%
It is assumed that $1-\kappa ^{2}\lambda >0$ and $\Omega ^{2}-k>0$ leading
to $r_{2}>0$.

The general solution to equation (\ref{eqnonlocal4}) is%
\begin{equation*}
U\left( \xi \right) =C_{1}\cosh \left( p_{1}\xi \right) +C_{2}\sinh \left(
p_{1}\xi \right) +C_{3}\cos \left( p_{2}\xi \right) +C_{4}\sin \left(
p_{2}\xi \right) ,
\end{equation*}%
with
\begin{equation}
p_{1}=\sqrt{\frac{-r_{1}+\sqrt{r_{1}^{2}+4r_{2}}}{2}},\;\;p_{2}=\sqrt{\frac{%
r_{1}+\sqrt{r_{1}^{2}+4r_{2}}}{2}}.  \label{ps}
\end{equation}%
Unknown constants $C_{1},...,C_{4}$ should be obtained using boundary
conditions (\ref{bc-nonloc1}) and (\ref{bc-nonloc2}). For the existence of
non-trivial solution of the homogeneous system with respect to constants, it
is required that%
\begin{eqnarray}
&&\sqrt{\text{$r_{2}$}}\left( \text{$r_{1}^{2}$}+2\text{$r_{2}$}+2\left(
\kappa ^{2}\text{$r_{2}$}\right) ^{2}-2\kappa ^{2}\text{$r_{1}r_{2}$}\right)
\notag \\
&&\qquad \qquad +\left( \text{$r_{2}$}+\kappa ^{2}\text{$r_{1}r_{2}$}-\left(
\kappa ^{2}\text{$r_{2}$}\right) ^{2}\right) \left( 2\sqrt{\text{$r_{2}$}}%
\cosh \text{$p_{1}$}\cos \text{$p_{2}$}+\text{$r_{1}$}\sinh \text{$p_{1}$}%
\sin \text{$p_{2}$}\right) =0.  \label{Det}
\end{eqnarray}

For the case when $\kappa =0$, relations (\ref{rs}), (\ref{ps}) and (\ref%
{Det}) reduce to
\begin{gather*}
r_{1}=\lambda ,\;\;r_{2}=\Omega ^{2}-k,\;\;p_{1}=\sqrt{\frac{-\lambda +\sqrt{%
\lambda ^{2}+4\left( \Omega ^{2}-k\right) }}{2}},\;\;p_{2}=\sqrt{\frac{%
\lambda +\sqrt{\lambda ^{2}+4\left( \Omega ^{2}-k\right) }}{2}}, \\
\lambda ^{2}+2\left( \Omega ^{2}-k\right) \left( 1+\cosh \text{$p_{1}$}\cos
\text{$p_{2}$}\right) +\lambda \sqrt{\Omega ^{2}-k}\sinh \text{$p_{1}$}\sin
\text{$p_{2}$}=0,
\end{gather*}%
giving the relations corresponding to Beck's column on Winkler foundation as
shown in \cite{KirSey} and to classical Beck's column if additionally $k=0$,
see \cite{a-ster}.

Equation (\ref{Det}) will be numerically solved in order to determine the
effect of non-locality parameter on dynamic stability boundary and to
determine if the Herrmann-Smith paradox is removed. This can be
mathematically stated as follows: For a given non-locality parameter $\kappa
$ and foundation stiffness $k$, determine load intensity $\lambda _{cr}$ and
frequency $\Omega _{cr}$ so that the first root of equation (\ref{Det}) has
multiplicity two. Numerical procedure of obtaining $\lambda _{cr}$ and $%
\Omega _{cr}$ was performed for equation (\ref{Det}) using various values of
$\kappa $ and $k$. The results are summarised in Table \ref{table1}.
\begin{table}[h]
\centering
\begin{tabular}{@{}c|c|c|c|c}
& \multicolumn{2}{c|}{$k=0$} & $k=5$ & $k=10$ \\ \hline
$\kappa $ & $\lambda _{cr}$ & $\Omega _{cr}$ & $\Omega _{cr}$ & $\Omega
_{cr} $ \\ \hline
$0$ & $20.05095$ & $11.01$ & $11.24$ & $11.46$ \\
$0.1$ & $16.79301$ & $9.32$ &  &  \\
$0.2$ & $11.24116$ & $6.35$ & $6.73$ & $7.09$ \\
$0.3$ & $7.22022$ & $4.13$ &  &  \\
$0.4$ & $4.80283$ & $2.77$ & $3.56$ & $4.20$ \\
$0.5$ & $3.35534$ & $1.94$ &  &  \\
$0.6$ & $2.45147$ & $1.42$ & $2.65$ & $3.47$%
\end{tabular}%
\caption{Critical load intensities $\protect\lambda _{cr}$ and frequencies $%
\Omega _{cr}$ for different values of non-locality parameter $\protect\kappa
$ and foundation stiffness $k$.}
\label{table1}
\end{table}

\begin{figure}[h]
\centering
\includegraphics[scale=0.85]{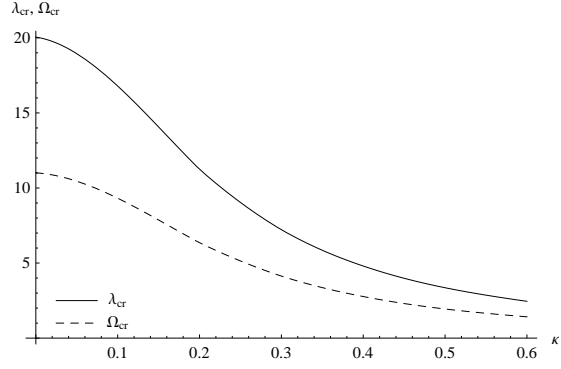}
\caption{Dependence of critical load $\protect\lambda _{cr}$ and frequency $%
\Omega _{cr}$ on non-locality parameter $\protect\kappa $.}
\label{Fignonlocal}
\end{figure}

For the case when Bernoulli-Euler moment-curvature relation is assumed ($%
\kappa $ and $k$ equal zero), the critical load $\lambda _{cr}=20.05095$ and
critical frequency $\Omega _{cr}=11.01$ corresponding to classical Beck's
column are reobtained, see \cite{a-ster}. This critical load remains
unchanged when foundations of varying stiffness are introduced thus
recovering the Herrmann-Smith paradox. The results also show that the
critical frequency increased with increasing foundation stiffness, which was
obtained in \cite{HerSmi}. By introducing non-local constitutive equation,
the critical load and frequency decrease with the increase of non-locality
parameter, as depicted in Figure \ref{Fignonlocal}.
This effect of non-locality parameter on stability boundary was also shown
to hold true for static problems of various conservative loading
configurations and boundary conditions in \cite{ANVZ,Ch-Wa,Wa-Zh-1}. Here it
is shown that reduction in critical load also occurs for non-conservative
problems when non-local constitutive equation is adopted. As with the
Bernoulli-Euler Beck's column on Winkler foundation, there was no change in
critical force for non zero values of foundation stiffness when adopting
non-local model. For a fixed value of non-locality parameter, the effect of
increasing the foundation stiffness on critical frequency was the same as
for Bernoulli-Euler rod. Thus the Herrmann-Smith paradox remains.

\section{Dynamic stability analysis for fractional viscoelastic rod}

The Herrmann-Smith and Ziegler paradoxes are now examined by determining
stability boundaries for Beck's column described using viscoelastic
constitutive equation of fractional Kelvin-Voigt type. Combining (\ref{DL1}%
), (\ref{DL2}), (\ref{DL4}) it is obtained that
\begin{equation}
\left( 1+a\,{}_{0}\mathrm{D}_{t}^{\alpha }\right) \frac{\partial ^{4}}{%
\partial \xi ^{4}}y\left( \xi ,t\right) +\lambda \frac{\partial ^{2}}{%
\partial \xi ^{2}}y\left( \xi ,t\right) +\frac{\partial ^{2}}{\partial t^{2}}%
y\left( \xi ,t\right) +ky\left( \xi ,t\right) =0,  \label{eqviscoelastic}
\end{equation}%
subject to boundary (\ref{DL5}) and initial (\ref{DL6}) conditions%
\begin{gather}
y(0,t)=0,\;\;\frac{\partial }{\partial \xi }y(0,t)=0,  \label{bc-ve1} \\
\left( 1+a\,{}_{0}\mathrm{D}_{t}^{\alpha }\right) \frac{\partial ^{2}}{%
\partial \xi ^{2}}y\left( 1,t\right) =0,\;\;\left( 1+a\,{}_{0}\mathrm{D}%
_{t}^{\alpha }\right) \frac{\partial ^{3}}{\partial \xi ^{3}}y\left(
1,t\right) =0,  \label{bc-ve2} \\
y\left( \xi ,0\right) =y_{0}\left( \xi \right) ,\;\;\frac{\partial }{%
\partial t}y\left( \xi ,0\right) =y_{1}\left( \xi \right) .  \label{bc-ve3}
\end{gather}%
Note that equation (\ref{eqviscoelastic}) reduces to the corresponding one
for elastic model when $a\rightarrow 0$ or $\alpha \rightarrow 0$. Contrary
to the approach used in Section \ref{nonlocalsec}, the Laplace transform
method will be implemented to analyse dynamic stability. Applying the
Laplace transform to (\ref{eqviscoelastic}) - (\ref{bc-ve2}) results in
\begin{gather}
\left( 1+as^{\alpha }\right) \frac{\partial ^{4}}{\partial \xi ^{4}}Y\left(
\xi ,s\right) +\lambda \frac{\partial ^{2}}{\partial \xi ^{2}}Y\left( \xi
,s\right) +(s^{2}+k)Y\left( \xi ,s\right) =sy_{0}\left( \xi \right)
+y_{1}\left( \xi \right) ,  \label{eqve1} \\
Y(0,s)=0,\;\;\frac{\partial }{\partial \xi }Y(0,s)=0,  \label{eqve2} \\
\left( 1+as^{\alpha }\right) \frac{\partial ^{2}}{\partial \xi ^{2}}Y\left(
1,s\right) =0,\;\;\left( 1+as^{\alpha }\right) \frac{\partial ^{3}}{\partial
\xi ^{3}}Y\left( 1,s\right) =0,  \label{eqve3}
\end{gather}%
where the Laplace transform of a function $f$ is defined by%
\begin{equation*}
F\left( s\right) =\mathcal{L}\left[ f\left( t\right) \right] \left( s\right)
=\int_{0}^{\infty }f\left( t\right) \mathrm{e}^{-st}\mathrm{d}t,
\end{equation*}%
and the Laplace transform of a Riemann-Liouville fractional derivative is
\begin{equation*}
\mathcal{L}\left[ {}_{0}\mathrm{D}_{t}^{\alpha }f(t)\right] \left( s\right)
=s^{\alpha }F(s)-\left[ \frac{1}{\Gamma (1-\alpha )}\int_{0}^{t}\frac{f(\tau
)}{(t-\tau )^{\alpha }}d\tau \right] _{t=0}=s^{\alpha }F(s),
\end{equation*}%
provided that $f$ is an exponentially bounded function, see \cite{TAFDE}.
Equation (\ref{eqve1}) reduces to%
\begin{equation}
\frac{\partial ^{4}}{\partial \xi ^{4}}Y\left( \xi ,s\right) +r_{1}(s)\frac{%
\partial ^{2}}{\partial \xi ^{2}}Y\left( \xi ,s\right) +r_{2}(s)Y\left( \xi
,s\right) =F(\xi ,s),  \label{eqve4}
\end{equation}%
with
\begin{equation}
r_{1}(s)=\frac{\lambda }{1+as^{\alpha }},\;\;r_{2}(s)=\frac{s^{2}+k}{%
1+as^{\alpha }},\;\;F(\xi ,s)=\frac{sy_{0}\left( \xi \right) +y_{1}\left(
\xi \right) }{1+as^{\alpha }}.  \label{eqve4-1}
\end{equation}%
The general solution to equation (\ref{eqve4}) is%
\begin{equation}
Y(\xi ,s)=Y_{H}(\xi ,s)+Y_{P}(\xi ,s),  \label{eqve5}
\end{equation}%
where
\begin{equation}
Y_{H}(\xi ,s)=C_{1}(s)\cosh \left( p_{1}(s)\xi \right) +C_{2}(s)\sinh \left(
p_{1}(s)\xi \right) +C_{3}(s)\cos \left( p_{2}(s)\xi \right) +C_{4}(s)\sin
\left( p_{2}(s)\xi \right) ,  \label{eqve5-1}
\end{equation}%
is the solution of homogeneous equation, with
\begin{equation*}
p_{1}=\sqrt{\frac{-r_{1}+\sqrt{r_{1}^{2}-4r_{2}}}{2}},\;\;p_{2}=\sqrt{\frac{%
r_{1}+\sqrt{r_{1}^{2}-4r_{2}}}{2}},
\end{equation*}%
and $Y_{P}$ is the particular solution of equation (\ref{eqve4}).

The general solution (\ref{eqve5}) takes the form
\begin{eqnarray}
Y(\xi ,s) &=&\frac{1}{D(s)}\left( D_{C_{1}}(s)\cosh \left( p_{1}(s)\xi
\right) +D_{C_{2}}(s)\sinh \left( p_{1}(s)\xi \right) \right.  \notag \\
&&+\left. D_{C_{3}}(s)\cos \left( p_{2}(s)\xi \right) +D_{C_{4}}(s)\sin
\left( p_{2}(s)\xi \right) \right) +Y_{P}(\xi ,s),  \label{eqve7}
\end{eqnarray}%
with the fact that $C_{1}=\frac{D_{C_{1}}}{D},...,C_{4}=\frac{D_{C_{4}}}{D},$
where $D,D_{C_{1}},...,D_{C_{4}}$ are the determinants corresponding to
system
\begin{eqnarray*}
Y(0,s)=0 &=&C_{1}(s)+C_{3}(s)+Y_{P}(0,s), \\
\frac{\partial }{\partial \xi }Y(0,s)=0 &=&C_{2}(s)p_{1}(s)+C_{4}(s)p_{2}(s)+%
\frac{\partial }{\partial \xi }Y_{P}(0,s), \\
\frac{\partial ^{2}}{\partial \xi ^{2}}Y(1,s)=0 &=&C_{1}(s)p_{1}^{2}(s)\cosh
\left( p_{1}(s)\right) +C_{2}(s)p_{1}^{2}(s)\sinh \left( p_{1}(s)\right) \\
&&-C_{3}(s)p_{2}^{2}(s)\cos (p_{2}(s))-C_{4}(s)p_{2}^{2}(s)\sin \left(
p_{2}(s)\right) +\frac{\partial ^{2}}{\partial \xi ^{2}}Y_{P}(1,s), \\
\frac{\partial ^{3}}{\partial \xi ^{3}}Y(1,s)=0 &=&C_{1}(s)p_{1}^{3}(s)\sinh
\left( p_{1}(s)\right) +C_{2}(s)p_{1}^{3}(s)\cosh \left( p_{1}(s)\right) \\
&&+C_{3}(s)p_{2}^{3}(s)\sin (p_{2}(s))-C_{4}(s)p_{2}^{3}(s)\cos \left(
p_{2}(s)\right) +\frac{\partial ^{3}}{\partial \xi ^{3}}Y_{P}(1,s).
\end{eqnarray*}%
This system is obtained by substitution of boundary conditions (\ref{eqve2})
and (\ref{eqve3}) in (\ref{eqve5}) and (\ref{eqve5-1}). The particular
solution $Y_{P}$ is dependent on initial conditions, see (\ref{eqve4}) and (%
\ref{eqve4-1}). Following the standard procedure for stability analysis, the
stability boundaries will be determined regardless of the choice of initial
conditions. Thus the determinant of system
\begin{equation}
D\left( s\right) =\text{$r_{1}^{2}\left( s\right) $}-2\text{$r_{2}\left(
s\right) $}\left( 1+\cosh \left( \text{$p_{1}\left( s\right) $}\right) \cos
\left( \text{$p_{2}\left( s\right) $}\right) \right) +\text{$r_{1}\left(
s\right) $}\sqrt{-\text{$r_{2}\left( s\right) $}}\text{$\sinh $}\left( \text{%
$p_{1}\left( s\right) $}\right) \sin \left( \text{$p_{2}\left( s\right) $}%
\right)  \label{eqve8}
\end{equation}%
will be the focus of stability analysis. The position of zeroes of (\ref%
{eqve8}) will determine the dynamic behaviour of the rod. If zeroes are
positioned in the left complex half plane or on the imaginary axis, the rod
will be stable with decreasing or constant amplitude of oscillation
respectively. Loss of stability occurs when zero has positive real part
resulting in vibrations of increasing amplitude. Thus the critical force is
determined as the force corresponding to zero of (\ref{eqve8}) having real
part equal to zero. The corresponding value of imaginary part represents the
critical frequency. For similar method of stability boundary analysis, we
refer to \cite{StAt1,StAt}. The critical force $\lambda _{cr}$ and frequency
$\Omega _{cr}$ will be numerically determined using (\ref{eqve8}) for
various values of order of fractional differentiation $\alpha $ and
foundation stiffness $k$.

Table \ref{table2} shows the values of critical force $\lambda _{cr}$ and
frequency $\Omega _{cr}$ for viscoelastic Beck's column with fixed value of
model parameter $a=0.4$ when order of differentiation $\alpha $ is varied.
\begin{table}[h]
\centering
\begin{tabular}{@{}c|c|c}
$\alpha $ & $\lambda _{cr}$ & $\Omega _{cr}$ \\ \hline
$0$ & $28.0713$ & $13.0431$ \\
$0.01$ & $15.4000$ & $6.4055$ \\
$0.1$ & $16.1912$ & $6.5619$ \\
$0.3$ & $18.5037$ & $6.9258$ \\
$0.5$ & $22.2605$ & $7.2553$ \\
$0.7$ & $29.2249$ & $7.4182$ \\
$0.9$ & $42.2860$ & $7.1619$ \\
$1$ & $51.4069$ & $6.7645$%
\end{tabular}%
\caption{Critical load intensities $\protect\lambda _{cr}$ and frequencies $%
\Omega _{cr}$ for different values of order of fractional differentiation $%
\protect\alpha $, with foundation stiffness $k=0,$ and model parameter $%
a=0.4 $.}
\label{table2}
\end{table}
For the case when $\alpha $ is zero, the dimensionless fractional
Kelvin-Voigt model (\ref{DL4}) reduces to Bernoulli-Euler equation $m=(1+a)%
\frac{\partial }{\partial \xi }\vartheta $. The critical force $\lambda
_{cr}=28.0713$ corresponding to this case reduces to the critical force $%
\lambda _{cr}=20.0510$ for classical Beck's column when divided by $1+a$. By
introducing small values of $\alpha $, the model becomes of fractional
Kelvin-Voigt type and there is a reduction in critical force. This result is
the Ziegler paradox. In \cite{Bolotin}, it was found that for extremely
small values of viscoelastic model parameter, thus approaching elastic
model, the critical force is less then in the elastic case (model parameter
is zero). The same result was achieved in our analysis however the order of
fractional differentiation approached zero, reducing the fractional
derivative of a function to a function itself, thus obtaining the elastic
model, as opposed to the case when the elastic model was recovered by
introduction of small model parameter.

\begin{figure}[h]
\centering
\includegraphics[scale=0.85]{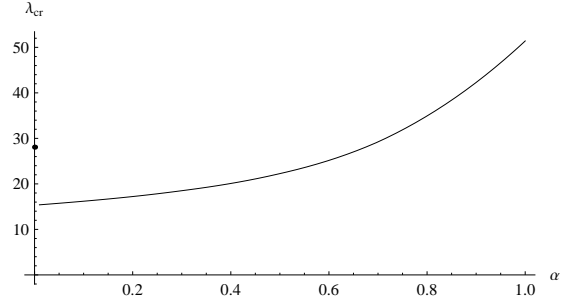}
\caption{Dependence of critical load $\protect\lambda _{cr}$ on order of
fractional differentiation $\protect\alpha $ with dot representing critical
load for elastic case.}
\label{Figviscoload}
\end{figure}

\begin{figure}[h]
\centering
\includegraphics[scale=0.85]{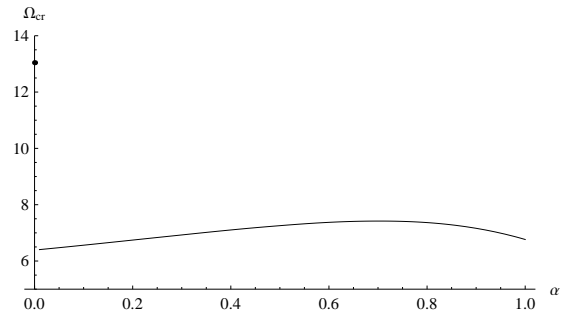}
\caption{Dependence of critical frequency $\Omega _{cr}$ on order of
fractional differentiation $\protect\alpha $ with dot representing critical
frequency for elastic case.}
\label{Figviscofreq}
\end{figure}

The critical force increases with the order of fractional differentiation,
see Figure \ref{Figviscoload}, as is to be expected due to greater
dissipation.
A decrease in the value of critical frequency occurs when small order of
differentiation is introduced, similarly to the behaviour of critical force.
The critical frequency however shows non-monotonic dependence on the order
of differentiation with maximum value occurring in the range of $\alpha \in
(0.5,0.9)$ as shown in Figure \ref{Figviscofreq}.

The effect of foundation stiffness on critical load and frequency is
analysed for given values of order of differentiation $\alpha =0.3$ and $%
\alpha =0.9$ with model parameter $a=0.4$. The results of this analysis are
shown in Table \ref{tabele3}.
\begin{table}[h]
\centering
\begin{tabular}{@{}c|c|c|c|c}
& \multicolumn{2}{c|}{$\alpha =0.3$} & \multicolumn{2}{c}{$\alpha =0.9$} \\
\hline
$k$ & $\lambda _{cr}$ & $\Omega _{cr}$ & $\lambda _{cr}$ & $\Omega _{cr}$ \\
\hline
$0$ & $18.5037$ & $6.9258$ & $42.2860$ & $7.1619$ \\
$5$ & $18.6295$ & $7.2985$ & $44.6815$ & $7.6008$ \\
$10$ & $18.7446$ & $7.6514$ & $46.9197$ & $8.0094$ \\
$20$ & $18.9495$ & $8.3086$ & $51.0257$ & $8.7559$ \\
$40$ & $19.2880$ & $9.4785$ & $58.1810$ & $10.0508$%
\end{tabular}%
\caption{Critical load intensities $\protect\lambda _{cr}$ and frequencies $%
\Omega _{cr}$ for different values of foundation stiffness $k$, and model
parameter $a=0.4$.}
\label{tabele3}
\end{table}
The presence of foundation is shown to influence the value of critical load
causing it to increase as the foundation stiffness does. The Herrmann-Smith
paradox is thus resolved when adopting fractional Kelvin-Voigt
moment-curvature constitutive equation. The critical frequency also
increases as in the case of classical Beck's column on elastic foundation.

\section{Conclusion}

The stability boundaries for Beck's column positioned on elastic foundation
have been analysed when adopting non-local and viscoelastic moment-curvature
constitutive equations for the column with the aim of removing the
Herrmann-Smith paradox. For the case of nano-rod of Eringen non-local type,
the separation of variables technique was adopted in order to determine
critical load causing dynamic instability, whereas the Laplace transform
method was utilised for the same analysis of viscoelastic rod of fractional
Kelvin-Voigt type.

The introduction of non-locality was found to reduce the load causing
flutter which is generally true for nano-rods. The Herrmann-Smith paradox
remained as the critical load is independent of foundation properties whilst
the critical frequency increased when the foundation stiffness did. The
removal of the Herrmann-Smith paradox was achieved by adoption of fractional
Kelvin-Voigt moment-curvature relation describing the column. In this case,
the critical load increased with higher orders of fractional differentiation
which is to be expected due to increased dissipation. However for orders of
fractional differentiation close to zero, the load causing dynamic
instability was less then for the elastic case despite the fact that the
viscoelastic model approached the elastic one. Therefore the Ziegler paradox
of destabilization is recovered.

\section*{Acknowledgement}

The authors would like to express great appreciation to Dr. Zora Vrcelj for
whom without this project would have not been possible and for her continual
support.

This research is supported by the Serbian Ministry of Education and Science
project $174005$, as well as by the Secretariat for Science of Vojvodina
project $114-451-3605/2013$.

Yanni Bouras acknowledges Victoria University for the financial support
provided for travel expenses.


\end{document}